\documentstyle[12pt]{article}
\newcommand\be{\begin{equation}}
\newcommand\ee{\end{equation}}
\newcommand\ba{\begin{eqnarray}}
\newcommand\ea{\end{eqnarray}}

\newcommand\un{\underline}

\newcommand\cP{\cal P}

\newcommand\bear{\begin{eqnarray*}}
\newcommand\eear{\end{eqnarray*}}

\begin{document}
\begin{titlepage}
\setlength{\textwidth}{5.0in}
\setlength{\textheight}{7.5in}
\setlength{\parskip}{0.0in}
\setlength{\baselineskip}{18.2pt}
\hfill
\setcounter{page}{1}
\begin{flushright}
HD--THEP--02--24\\
\end{flushright}
\vskip1.5cm
\begin{center}
{\large{\bf On the Hamilton-Jacobi equation for second class constrained systems}}
\\
\vspace{1cm}
K.D. Rothe 
\footnote{email: k.rothe@thphys.uni-heidelberg.de}
\\
{\it Institut  f\"ur Theoretische Physik - Universit\"at Heidelberg}\\
{\it Philosophenweg 16, D-69120 Heidelberg, Germany}
\\
and
\\
F.G. Scholtz 
\footnote{email: fgs@sun.ac.za}
\\
{\it Institute of Theoretical Physics, University of Stellenbosch, Stellenbosch 7600, South Africa}

{(April 2003)}
\end{center}

\begin{abstract}
\noindent
We discuss a general procedure for arriving at the Hamilton-Jacobi equation
of second-class constrained systems, and illustrate it in terms of a number of examples by explicitely obtaining the respective Hamilton principal function, and verifying that it leads to the correct solution to the Euler-Lagrange equations.

\end{abstract}
\end{titlepage}

\newpage

\section{Introduction}

There has recently been much interest in a Hamilton-Jacobi (HJ) formulation 
for constrained systems \cite{Guler92a}-\cite{baleanu}. Although there exists consensus in 
the formulation for purely first class systems, the case of second-class systems
has to be dealt with separately. 

For an unconstrained system described by a Lagrangian $L(q_i,\dot q_i)$, $i=1,\cdots,n$, the HJ-equations read,
\be\label{HJ-unconstrained}
H_0' \equiv p_0 + H_0(q_0,q_i,p_i) = 0
\ee
where $H_0$ is the canonical Hamiltonian, and where, for later convenience,
$p_0$ and $p_i$ are shorthand for the partial derivatives
\be\label{partialS}
p_{0}=\frac{\partial S}{\partial q_0}\,,\quad
p_{i}=\frac{\partial S}{\partial q_{i}}\,,\quad q_0 = t\,,
\ee
with $S$ the Hamilton principal function $S = S(t,q_i)$ to be obtained as
a solution of eq. (\ref{HJ-unconstrained}).

Consider now a constrained system in which not all canonical 
variables are independent.  In that case, the Lagrangian 
$L(q_{i},\dot{q}_{i})$ is singular, so that the determinant of 
the Hessian matrix $H_{ij}=\frac{\partial^2 L}
{\partial\dot{q}_i\partial\dot{q}_j}$ is zero 
and the accelerations of some variables ${q}_i$ are not 
uniquely determined by the positions and the velocities at a given time. 

Let $n-m_1$ be the rank of the Hessian.  In that case only $n-m_1$ velocities 
$\dot{q}_a$ ($a=m_1+1,...,n$) can be solved for as a function of the coordinates $q_i$, the  momenta $p_a$ canonically conjugate to $q_{a}$, and the remaining
velocities $\dot q_\alpha$, to yield 
$\dot{q}_a=\dot{q}_a (q_i,p_b,\dot q_\beta)$.  The remaining momenta $p_\alpha$ 
($\alpha=1,...,m_1$) can be shown to be functions of $q_i$ 
and $p_a$ only, and represent the primary constraints 
$\phi_\alpha(q,p)\equiv p_\alpha + H_\alpha(q_i,p_a) = 0$,
in the Dirac terminology~\cite{Dirac}. We write them in a form appropriate for our purposes:
\be\label{primary}
H'_\alpha \equiv p_\alpha 
+ H_\alpha(q_0,q_i,p_a)=0\,,\quad (\alpha = 1,\cdots m_1) 
\ee
Adjoining these constraints to (\ref{HJ-unconstrained}), we are led to
consider the coupled set of differential equations
\ba\label{HJ-constrained1}
&&H'_0 \equiv \frac{\partial S}{\partial q_0} + 
H_0\left(q_0,q_\alpha,q_a,
\frac{\partial S}{\partial q_a}\right)=0\,,\nonumber\\
&&H'_{\alpha} \equiv \frac{\partial S}{\partial q_{\alpha}} + 
H_{\alpha}\left(q_0,q_\alpha,q_a,\frac{\partial S}{\partial q_{a}}\right)
= 0\,,\quad {\alpha}=1,...,m_1\,,
\ea
where $H_0$ is the canonical Hamiltonian evaluated on the subspace of the
{\it primary} constraints, which only depends on the
variables $q_i$, $p_a$ (and possibly $q_0 = t$), as is well known.   
Continuing the above line of reasoning, the system of differential equations (\ref{HJ-constrained1}) should further be suplemented with the corresponding
set of equations associated with possible secondary constraints
$\varphi(q,p)_\sigma=0$. Let there be $m_2$ such constraints. Adjoining these to (\ref{HJ-constrained1}), we have in our generalized notation,
\be\label{HJ-constrained2}
H'_{\un\alpha}({q_{\un i}},\frac{\partial S}{\partial q_{\un i}})=0\,,\quad 
\un i = 0,1,\cdots, n\,,\quad \un\alpha = 0,1,\cdots, m= m_1 + m_2 \,.
\ee
Assuming that the secondary constraints do not represent constraints
among the coordinates $q_{\un i}$ themselves,
\footnote{As we shall see, this condition is not satisfied for the
``non-linear sigma model".}
this coupled set of differential equations only admits a solution provided
the $H'_{\un\alpha}$ are in weak involution in the sense
\footnote{On the space of solutions to (\ref{HJ-constrained1}) this then
implies strong involution \cite{Caratheodory}.}
\be\label{integrability}
\{H'_{\un\alpha},H'_{\un\beta}\}|_{p_{\un i}=\frac{\partial S}{\partial q_{\un i}}} 
= C_{\un\alpha,\un\beta}^{\,\,\,\,\,\,\,\un\gamma}H'_{\un\gamma}(q_{\un i},\frac{\partial S}{\partial q_{\un i}})\,,
\ee
where the (generalized) Poisson bracket is defined by
\be\label{genPoisson}
\{A,B\} = \sum_{\un i=0}^{m}\left(\frac{\partial A}{\partial q_{\un i}}
\frac{\partial B}{\partial p_{\un i}} -  
\frac{\partial B}{\partial q_{\un i}}\frac{\partial A}{\partial p_{\un i}}\right)\,.
\ee
Hence, for second class systems, {\it  our Ansatz (\ref{HJ-constrained2}) for the HJ equations must be incorrect}. 

One way of circumventing the integrability problem would be to embed the model in question into a
larger phase space, such as to become first class \cite{HR}. Then the integrability
conditions (\ref{integrability}) are statisfied and our Ansatz
is self-consistent \cite{Caratheodory}. This is, however, not the point of view we wish to take. As we shall rather illustrate in terms of a number of examples,
a direct way of proceeding in the case of systems with second class constraints is to perform a canonical change of variables in which the second-class constraints become part of the variables. If the Poisson brackets of constraints are given by constant matrices, one can always perform a transformation to variables, where
the constraints themselves can be grouped into canonically conjugate pairs.
Otherwise this may not be possible \cite{Senja}, and the embedding procedure would provide a way out of the dilemma.

Unlike the discussions found in the literature
\cite{Guler92a,Guler92b,Guler92c} it is the final objective of this paper to actually construct
the Hamilton-principal function $S$ for the examples we consider, and to obtain from there the solution of the corresponding Euler-Lagrange equations
following standard methods.

Our strategy to be illustrated by a number of examples will be the following:
We begin by seeking all constraints, following the Dirac algorithm.
We then seek a canonical transformation from a set of phase space variables $q_i,p_i$ 
to a new set $(Q,P)$ = ($q^*_{a},\chi_\alpha,p^*_{a},{\cP}_\alpha$), in which the complete set of primary and secondary constraints 
are grouped into canonically conjugate pairs ($\chi_\alpha,{\cP}_\alpha$)
satisfying $\{\chi_\alpha,{\cP}_\beta\} = \delta_{\alpha\beta}$.
Since the number of second class constraints is even, this is
always possible for linear constraints.
\footnote{As shown in ref. \cite{Senja} this is not always possible, if the constraints are nonlinear.}
The remaining set of variables 
$(q_{a}^*,p_{a}^*)$, which in turn are canonically conjugate to each other, $\{q_{a}^*,p_{b}^*\} = \delta_{ab}$, are chosen to commute with 
the set ($\chi_\alpha,{\cP}_\alpha$).
Denoting by $F_2(q;p^*,{\cP} ;t)$ the generating
function for this canonical transformation, the new Hamiltonian
will be given by $\tilde H(q^*,p^*;\chi,{\cP} ;t)
 = H(q(q^*,p^*,\chi,{\cP} ),p(q^*,p^*,\chi,{\cP});t)
+ \frac{\partial F_2}{\partial t}$, where $q(q^*,p^*,\chi,{\cP})$ and $p(q^*,p^*,\chi,{\cP})$ are obtained by solving the coupled set of equations
$p_i = \frac{\partial F_2}{\partial q_i}$, 
$\chi_\alpha = \frac{\partial F_2}{\partial {\cP}_\alpha}$ and 
$q_a^* = \frac{\partial F_2}{\partial p_a^*}$.
The extended action in terms of the new variables now reads,
\footnote{To formulate the general problem it is convenient to work
with the extended action, where all of the constraints are implemented
via Lagrange multipliers. Since all the constraints are supposed
to be generated from the total Hamiltonian $H_T$, involving only
the primary constraints, via the Dirac self-consistent
algorithm, the Lagrange multipliers associated with the secondary constraints
will eventually turn out to be zero in a variational calculation.}
\be\label{extended-action}
S_E = \int dt\{p^*_{a}\dot q^*_{a} + {\cP}_\alpha \dot\chi_\alpha
- \tilde H(q^*,p^*,\chi,{\cP} ;t) - 
\lambda_\alpha\chi_\alpha - \eta_\alpha {\cP}_\alpha\}
\ee
Since the constraints commute with 
$q_{a}^*$ and $p_{a}^*$, the equations of motion for these variables
are determined by $\tilde H$ with the constraints set equal to zero, strongly:
\[
\dot q_{a}^* = \{q_{a}^*,\tilde H(q^*,p^*,0,0;t)\}\,,\quad
\dot p_{a}^* = \{p_{a}^*,\tilde H(q^*,p^*,0,0;t)\}
\]
while the equations for the remaining coordinates, the constraints,
just reproduce the persistency equations, and represent a 
set of differential equations in the new coordinates. With respect
to the coordinates $(q^*,p^*)$ the problem has thus been reduced to that
of an unconstrained system, and the usual Hamilton-Jacobi treatment 
applies. Note that the canonical change in phase space variables has
now removed the inconsisteny of the originally envisaged HJ formulation.
The reason is that, unlike the Hamiltonian, the Hamilton Principal function
in the new coordinates $(Q,P)$ cannot be obtained by a simple substitution
$(q,p)\to (q(Q,P),p(Q,P))$.

In different terms: from the total action (\ref{extended-action})
we obtain for the HJ equation
\be\label{HJequation}
\tilde H(q^*,\frac{\partial S}{\partial q^*},\chi,\frac{\partial S}{\partial \chi},t)
+ \frac{\partial S}{\partial t} = 0
\ee
together with the constraint equations,
\be\label{constraints}
\chi_\alpha = 0\,,\quad {\cP}_\alpha = \frac{\partial S}{\partial {\chi}_\alpha}=0\,.
\ee
In general our problem thus reduces to solving the differential equations
\be\label{generalHJequations}
\hat H(q^*,\frac{\partial S}{\partial q^*},t) + \frac{\partial S}{\partial t} = 0
\ee
with
\be\label{Hhat}
\hat H(q^*,p^*) = \tilde H(q^*,p^*,\chi={\cP}=0)\,.
\ee

In the following section we illustrate these ideas in terms of a number of simple examples. We shall be rather detailed in the first example, in order to
illustrate various aspects of the problem.

\section{Second class systems and HJ formulation}

In this section we shall consider several simple quantum mechanical models of
purely second class systems and show how to properly deal with them in order
to construct the corresponding Hamilton Principal function $S$. We also verify
that the formalism yields the solutions to the corresponding equations of
motion.

\newpage

\bigskip\noindent
{\bf Example 1}

\bigskip
Consider the Lagrangean
\be\label{L1}
L = \frac{1}{2}\dot x^2 + \dot x y - \frac{1}{2}(x-y)^2
\ee
The corresponding Euler-Lagrange equations can be written in the form,
$\ddot x = 0\,\,\,, \dot x + x = y $.
We have one primary constraint
$\phi := p_y =0$, 
and the canonical Hamiltonian on the primary surface reads,
\be\label{H1}
H_0 = \frac{1}{2}(p_x - y)^2 + \frac{1}{2}(x - y)^2\,.
\ee
The usual Dirac algorithm yields a secondary constraint $\varphi = 0$,
\be\label{secondary1}
\{\phi,H_0\} = -2\varphi\,,\quad \varphi = y - \frac{1}{2}(p_x + x)\,.
\ee
There are no further constraints. Since $\{\varphi,\phi\} = 1$
these constraints are second class and canonically conjugate to each other.

If we were to proceed naivly and make the replacements $p_y = \frac{\partial S}{\partial y}$ in the primary
constraint $p_y=0$, then this would imply that $S$ is independent of $y$, which would be in direct conflict
with the equation obtained by making the substitution $p_x = \frac{\partial S}{\partial x}$ in the constraint $\varphi = 0$.
This reflects the non-commutative structure of our second class system, as well as the violation of the
integrability condition (\ref{integrability}).

In order to circumvent this difficulty, we make a canonical transformation to a new set of variables  $(q^*,p^*)$ and
$(\chi,{\cP})$, in which one of the canonically conjugate pairs
are chosen to be the constraints themselves:
\ba\label{newvariables1}
\chi &=& \varphi\,,\quad {\cP} = \phi\,,\nonumber\\
q^* &=& x - \frac{1}{2}p_y\,,\quad p^* = p_x + \frac{1}{2}p_y\,.
\ea
We evidently have
\be\label{newPB1}
\{\chi,{\cP}\} = 1\,,\quad  \{q^*, p^*\} = 1\,.
\ee
Futhermore,  $(q^*,p^*)$ and $(\chi,{\cP})$ ``commute". Our change of variables is thus canonical. Indeed one readily checks that
\[
p_x \dot x + p_y \dot y =
p^*\dot q^* + {\cP}\dot{\chi} + \frac{1}{2}\frac{d}{dt}(p^*{\cP} 
- \frac{1}{2}{\cP}^2)\,.
\]
Furthermore, the corresponding generating functional $F_2(x,y;p^*,{\cP})$ for the transformation
(\ref{newvariables1}) is readily constructed:
\[
F_2 = (p^* - \frac{1}{2}{\cP})x + {\cP} y -\frac{1}{2}{\cP}p^*
+ \frac{1}{8}{\cP}^2\,.
\]
The inverse transformation to (\ref{newvariables1}) reads
\ba\label{inversenewvariables1}
x &=& q^* + \frac{1}{2}{\cP} \,,\quad p_x = p^* - \frac{1}{2}{\cP}\,,\nonumber\\
y &=& \frac{1}{2}(q^* + p^*) + \chi\,,\quad p_y = {\cP} \,.
\ea
In terms of the new variables,
\be
\tilde H 
= \frac{1}{2}\left[\frac{1}{2}(p^*-q^*)-(\frac{1}{2}{\cP} + \chi)\right]^2
+ \frac{1}{2}\left[\frac{1}{2}(q^*-p^*)-(\frac{1}{2}{\cP}-\chi)\right]^2
\ee
One checks,
\ba
\dot q^* &=& \frac{1}{2}(p^* - q^*) - \frac{1}{2}{\cP}\nonumber\\
\dot p^* &=& \frac{1}{2}(p^* - q^*) - \frac{1}{2}{\cP}\nonumber\\
\dot{\cP} &=& -2\chi\nonumber\\
\dot{\chi} &=& \eta - \frac{1}{2}(p^*-q^*) + \frac{1}{2}{\cP}\nonumber\\
\ea
The 3'rd equation reproduces the secondary constraint, and the 4'th equation fixes the parameter $\eta$ as expected.

The above shows, that as a consequence of having the conjugate pairs in strong involution, we can simply set the constraints in $\tilde H_0$ equal to zero and
proceed with the Hamiltonian
\be\label{newH1}
\hat H_0 = \frac{1}{4}(p^*-q^*)^2\,.
\ee
Hence we are left with only one HJ equation,
\be\label{HJ1}
\frac{\partial S}{\partial t} 
+ \frac{1}{4}(\frac{\partial S}{\partial q^*} - q^*)^2 = 0
\ee
showing that $S$ will be just a function of $q^*$.
The HJ equation associated with the primary constraint ${\cP} = 0$ tells us that
$S$ is independent of the new coordinate $\chi$:
${\cP} = \frac{\partial S}{\partial \chi} = 0$.

Since the Hamiltonian does not depend explicitely on time, we must have
\[
\frac{\partial S}{\partial t} = -\alpha^2 = constant
\]
so that equation (\ref{HJ1}) has the solution
\be\label{S1}
S(q^*,\alpha,t) = \pm 2\alpha q^* + \frac{1}{2}q^{*2} - \alpha^2 t\,.
\ee
Notice that $S$ depends only on one integration constant, since when regarded as the generating function
taking us to new momenta $p^*$ and ${\cP}$, one of the momenta (${\cP}$) is just the primary constraint ${\cP} = \phi = 0$.
There is another constant arising from the equation
\[
\beta = \frac{\partial S}{\partial\alpha}\,,
\]
which implies from (\ref{S1}),
\[
\beta = \pm 2q^* - 2\alpha t\,.
\]
Together with (we choose the plus sign in (\ref{S1}), without loss of generality)
\[
p^* = \frac{\partial S}{\partial q^*} = 2\alpha + q^*
\]
we now solve the problem in the standard way. Taking account of the constraints
$\chi_\alpha = {\cP}_\alpha = 0$ in (\ref{inversenewvariables1}), we find
\[
x = \frac{1}{2}\beta + \alpha t\,,\quad y = (\alpha + \frac{1}{2}\beta) 
+ \alpha t
\]
which evidently is a solution of the Euler-Lagrange equations associated with (\ref{L1}).

\bigskip\noindent
{\bf Example 2: QM $\sigma$ - model}

\bigskip
Consider now the quantum mechanical version of the ``sigma model" as given by the Lagrangean
\be\label{L2}
L=\frac{\dot{\vec q}^2}{2}+\lambda(\vec{q}^2-1)
\ee
implying the primary constraint $p_\lambda=0$.
The canonical Hamiltonian on the primary surface reads,
\be\label{H2canonical}
H_0=\frac{\vec p^2}{2}-\lambda(\vec q^2-1)\,.
\ee
The complete set of primary and secondary constraints can be written in the form
\be\label{constraints2}
\phi=p_\lambda,\quad \varphi=\lambda+\frac{\vec p^2}{2}\,,\quad 
\varphi_1=\vec q^2-1\,,\quad \varphi_2=\frac{\vec p\cdot\vec q}{2\vec q^2}\,.
\ee
The Poisson brackets of the constraints show that they can be grouped into
mutually commuting, canonically conjugate pairs: 
\be\label{PB2}
\{\varphi,\phi\}=1,\quad \{\varphi_1,\varphi_2\}=1\,.
\ee
Let us consider for simplicity the case of two dimensions.
One may be tempted to make a transformation from $(\lambda,p_\lambda)$ to $(\varphi,\phi)$ and $(q_1,q_2,p_1,p_2)$ 
to $(\varphi_1,\varphi_2,\theta,J)$, where
($\theta,J$) are the canonically conjugate variables, 
\be\label{newvariables2}
\theta=\arctan(q_2/q_1)\,,\quad J=\epsilon_{ij}q_ip_j\,,
\ee
with $J$ evidently playing the role of the ``third" component of angular momentum.
The second set of variables commute with the first set, except for $\theta$, which does not commute with $\varphi$: $\{\varphi,\theta\}=J/\vec q^2$. Hence this transformation is not canonical.

In order to remove this problem we eliminate $\lambda$ and 
$p_\lambda$ using the constraints $\varphi=0$ and $\phi=p_\lambda=0$.
\footnote{It is easy to see, that this is a legitimate procedure
since $\lambda$ plays the role of a Lagrange multiplier in the Lagrangean.
More precisely, implementing strongly the constraint  $\varphi = 0$
in the Hamiltonean (\ref{H2canonical}) from the outset, leads to an
equivalent set of equations of motion.}
The resulting Hamiltonian is
$H_0=\frac{1}{2}\vec p^2 q^2$.
Noting that
\[
J^2 = \vec q^2\vec p^2 - (\vec q\cdot\vec p)^2 = \vec p^2 (1+\varphi_1) - 4\varphi_2^2 (\varphi_1 + 1)^2
\]
we then obtain for $\hat H_0$,
\[
\hat H_0 = \frac{1}{2}J^2
\]
We are thus left to solve the equation
\be\label{HJ2}
\frac{\partial S}{\partial t} 
+\frac{1}{2}\left(\frac{\partial S}{\partial\theta}\right)^2 = 0
\ee
which has the solution (an irrelevant additive constant has been ommitted)
\be\label{S2}
S(\theta,\alpha,t) = -\frac{\alpha^2}{2} t + \alpha\theta\,.
\ee
In order to obtain the solution to the Euler-Lagrange equations we recall
that we still have the equation
$\frac{\partial S}{\partial\alpha} = \beta$, which gives the solution in terms of two integrations constants:
\be\label{solution2}
\theta = \alpha t + \beta
\ee
One again verifies that this provides the solution of the 
Euler-Lagrange equations.

\bigskip\noindent
{\bf Example 3: Multidimensional rotator} 

\bigskip
Consider the Lagrangean \cite{HR,Guler01},
\[
L = \frac{1}{2}\dot{\vec q}^2 + \lambda \vec q\cdot\dot{\vec q}\,.
\]
There is one primary constraint
$\phi = p_\lambda = 0$
and the canonical Hamiltonian evaluated on the primary surface reads,
\be\label{H3canonical}
H_0 = \frac{1}{2}(\vec p - \lambda\vec q)^2\,.
\ee
The model effectively describes the motion on a $n-1$ dimensional sphere without specification of the
radius of the sphere. Note that unlike in the case of the $\sigma$-model, the momentum conjugate to
$\vec q$ is no longer the mecanical momentum: $\vec p = \dot{\vec q} + \lambda\vec q$. There is only one
secondary constraint
\be\label{S3-constraint}
\varphi = \lambda - \frac{\vec p\cdot\vec q}{\vec q^2}\,,
\ee
which has been chosen to be canonically conjugate to $\phi$:
$\{\varphi,\phi\} = 1$.

i) We first consider this model in two dimensions.
A coordinate change $(\lambda,p_\lambda)\to (\varphi,\phi)$ and $(q_1,p_1,q_2,p_2)\to (r,p_r,\theta,J)$
would again lead to the second set not commuting with the first, so that $\lambda$ should first be eliminated
using the constraint $\varphi = 0$. Having done this we may replace the canonical Hamiltonian (\ref{H3canonical}) by (see footnote 7)
\be\label{H3}
\hat H_0 = \frac{1}{2}\left(\vec p 
- \frac{(\vec p\cdot\vec q)}{\vec q^2}\vec q\right)^2 = \frac{J^2}{2r^2}
\ee
where $r=\sqrt{{\vec q}^2}$, $J$ is given in (\ref{newvariables2}), and $\{\theta,J\}=1$. 
Notice that $r$ just enters as a parameter. Replacing $J$ by $\left(\frac{\partial S}{\partial\theta}\right)$, we obtain from
(\ref{H3}) the HJ equation,
\be\label{HJ3}
\frac{\partial S}{\partial t} +\frac{1}{2r^2}\left(\frac{\partial S}{\partial\theta}\right)^2  = 0\,,
\ee
which just differs from (\ref{HJ2}) by $r$ being arbitrary, and not equal to one. Correspondingly (\ref{S2}) is replaced by
\be\label{S3}
S(\theta,\alpha,t) = -\frac{\alpha^2}{2} t + r\alpha\theta \,.
\ee
The rest proceeds as in the case of model 2.

ii) We now reconsider the problem in $n$-dimensions. Solving again the constraint $\varphi = 0$ for $\lambda$, the HJ equation (\ref{HJ3})
is replaced by,
\footnote{Note that the coordinates $q_a$ play the role of $q_a^*$ in section 1.}
\[
\frac{\partial S}{\partial t} + \frac{1}{2}\left(\frac{\partial S}{\partial q_a}\right)^2 
- \frac{1}{2}\left(\frac{\partial S}{\partial q_a}\right)
\left(\frac{\partial S}{\partial q_b}\right)\frac{q_aq_b}{\vec q^2} = 0\,.
\]
We make the Ansatz,
\be\label{Ansatz}
S = f(\vec n\cdot\vec q)
\ee
with $\vec n$ a unit normal vector parametrized by $n-1$ constants. The HJ equation for $f$ then reads,
\[
\frac{1}{2}\left(1 - \frac{(\vec n\cdot\vec q)^2}{\vec q^2}\right)
f'^2(\vec n\cdot\vec q) = \frac{\alpha^2}{2}\,.
\]
Setting $x = \vec n\cdot\vec q$ and $r^2 = \vec q^2$,  we then find
\[
f'(x) = \pm\frac{\alpha}{\sqrt{1 - \frac{x^2}{r^2}}}
\]
so that upon integration in $x$ the Hamilton principal function takes the form
\be\label{S4}
S = \alpha r \tan^{-1}\frac{\vec n\cdot\vec q}{\sqrt{r^2 - (\vec n\cdot\vec q)^2}} - \frac{\alpha^2}{2} t + const
\ee
The Hamilton principal function contains $n$ independent constants, which we take to be $\alpha$ and $n_1,n_2,\ldots n_{n-1}$, while the normalization of $\vec n$ implies for the $n$'th component, $n_n=\sqrt{1-\sum_{a=1}^{n-1}n_a^2}$.  Differentiating the Hamilton principal function with respect to these constants (new momenta in the corresponding generating functional) yields the $n$ time-independent new coordinates:
\be\label{constantcoordinates}
\beta=\frac{\partial S}{\partial\alpha}\,,\quad
\beta_a=\frac{\partial S}{\partial n_a}\,,\quad a=1,2,\ldots n-1\,.
\ee
From the first equation and (\ref{S4}) we easily obtain
\be\label{ndotq}
\vec n\cdot\vec q=r\cos\frac{\beta+\alpha t}{r}\equiv r\cos\Omega(t)\,.
\ee
Using this, we obtain from the second equation in (\ref{constantcoordinates})
\be\label{q-ccordinate}
q_a=\frac{\beta_a}{\alpha}\sin\Omega(t)+\frac{n_a q_n}{n_n};\quad a=1,2,\ldots,n-1
\ee
Multiplying this equation by $n_a$, summing from $a=1$ to $a=n-1$ and using the normalization of $\vec n$ we obtain from (\ref{q-ccordinate}),
\[
\vec n\cdot\vec q=\frac{1}{\alpha}\sum_{a=1}^{n-1}n_a\beta_a\sin\Omega(t)+\frac{q_n}{n_n}=r\cos\Omega(t)
\]
This allows us to solve for $q_n$ as 
\[
q_n=n_n\left(r\cos\Omega(t)-\frac{1}{\alpha}\sum_{a=1}^{n-1}n_a\beta_a\sin\Omega(t)\right)\,.
\]
Substituting back in the equation for $q_a$ we obtain
\[
q_a=\frac{1}{\alpha}\left(\beta_a-n_a\sum_{a=1}^{n-1}n_a\beta_a\right)\sin\Omega(t)+n_ar\cos\Omega(t)\,,\quad a=1,2,\ldots,n-1\,.
\]
Introducing the $n$-dimensional vector $\vec\beta=(\beta_1,\beta_2,\ldots,\beta_{n-1},0)$, these results can be written compactly as
\[
\vec q=\frac{1}{\alpha}\left(\vec\beta-(\vec n\cdot\vec\beta)\vec n\right)\sin\Omega(t)+r\vec n\cos\Omega(t)
\]
Substituting the above result into our original condition, $r^2=\vec q\,^2$, leads to 
\be\label{r}
r=\sqrt{\frac{\vec\beta^2-(\vec n\cdot\vec\beta)^2}{\alpha^2}}
\ee
Thus the radius of motion is fixed if the new time independent coordinates are specified.  More conventionally one would consider the radius of motion as a given initial condition; then the above condition eliminates one of the $\beta_a$ as an independent constant.  Indeed, from (\ref{r}) we see that the component of $\vec\beta$ orthogonal to $\vec n$ is fixed to be $\alpha r$. In this case one therefore considers $\alpha$, $r$ and the remaining $n-2$ constants, which determine the component of $\vec\beta$ parallel to $\vec n$ as independent constants.

\bigskip\noindent
{\bf Example 4: Landau model in the zero mass limit}

\bigskip
The Lagrangean of a spinless charged particle moving on a two dimensional plane
in a constant background magnetic field $B$ perpendicular to the 12-plane, and
a harmonic oscillator potential, can be written in the form
\[
L_{\it Landau} = \frac{m}{2}{\dot{\vec q}}^2 + \frac{B}{2}\vec q\times \dot{\vec q}
- \frac{k}{2}\vec q^2\,.
\]
In the zero-mass limit the Lagrangian of the model reduces to 
\be\label{L5}
L=\frac{B}{2}\vec q\times\dot{\vec q}-\frac{k}{2}\vec q^2
\ee
There are only two primary constraints
\[\phi_i=\frac{1}{\sqrt{B}}p_i+\frac{\sqrt{B}}{2}\epsilon_{ij}q_j
\]
with non-vanishing Poisson brackets
\[
\{\phi_i,\phi_j\}=\epsilon_{ij}
\]
We perform the following change of variables from $q_i,p_i$ to
$\phi_1,\phi_2,q,p$:
\footnote{In the notation of section 1, $q,p$ now play the role of $q^*,p^*$, and $\phi_1 = \chi, \phi_2 = {\cP}$.}
\ba\label{newvariables5}
\phi_1&=&\frac{1}{\sqrt{B}}p_1+\frac{\sqrt{B}}{2}q_2\,,\quad
\phi_2=\frac{1}{\sqrt{B}}p_2-\frac{\sqrt{B}}{2}q_1\,,\nonumber\\
q&=&\frac{1}{2}q_1+\frac{1}{B}p_2\,,\quad p=p_1-\frac{B}{2}q_2\,,
\ea
with the inverse transformation
\ba\label{inverse5}
q_1&=&q-\frac{1}{\sqrt{B}}\phi_2\,,\quad 
q_2=\frac{1}{B}(\sqrt{B}\phi_1-p)\,,\nonumber\\
p_1&=&\frac{1}{2}(\sqrt{B}\phi_1+p)\,,\quad 
p_2=\frac{1}{2}(Bq+\sqrt{B}\phi_2)\,.
\ea
In terms of the new coordinates the canonical Hamiltonian evaluated on the constraint surface reads,
\be\label{H5}
\hat H_0 = \frac{k}{2}\left(q^2 + \frac{1}{B^2}p^2\right)
\ee
Hence we have for the HJ equation
\be\label{HJ5}
\frac{\partial S}{\partial t} + \frac{k}{2B^2}\left(B^2q^2 + \left(\frac{\partial S}{\partial q}\right)\right) = 0\,,
\ee
having the solution
\be\label{S5}
S(q,\alpha,t) = -\alpha t + B\int^q dq'\,\sqrt{\frac{2\alpha}{k} - q'^2} + const.
\ee
From the equation $\beta = \frac{\partial S}{\partial\alpha}$ we then
obtain in the usual way the solution
\[
q(t) = \sqrt{\frac{2\alpha}{k}}\cos(\omega t + b)\,,
\]
with
$\omega ={k/B}$ and b = ${\beta k}/{B}$.

\bigskip\noindent
\section{Conclusion}

\bigskip
A {\it naive} extension of Hamilton-Jacobi (HJ) theory to second class constrained systems
is condemned to failure right from the outset since it leads to differential
equations which are in direct conflict. Mathematically this conflict is
expressed by the violation of the integrability condition (\ref{integrability}).

It is useful to reformulate these integrability conditions in terms of the
commutator algebra of operators. 
To this end we begin by introducing for each $H'_{\un\alpha}$ a linear differential operator $X_{\un{\alpha}}$ 
defined by  \cite{Caratheodory,Guler92b,Hong}
\begin{equation}
\label{partial}
X_{\un{\alpha}} f= \{ H^{\prime}_{\un{\alpha}},f \}\,,
\end{equation}
where the Poisson bracket is the generalized Poisson bracket defined in
(\ref{genPoisson}), and $f$ an arbitrary functions of 
$q_{\un i}$ and $p_{\un i}$. If $H_{\un\alpha}^{\prime}$ does
not possess an explicit time dependence, we recover the usual definition of 
the Poisson bracket.  Making use of the Jacobi identity for Poisson brackets, we obtain from  Eq.(\ref{partial})  the following 
relation between the commutator of the operators $X_{\un\alpha}$,
and the Poisson algebra of the constraints :
\begin{equation}\label{X-algebra}
[X_{\un{\alpha}}, X_{\un{\beta}}]f=\{\{H^{\prime}_{\un{\alpha}},
H^{\prime}_{\un{\beta}}\},f\}.
\end{equation}
for arbitrary $f$.  It is now clear that the commutator algebra
of the operators $X_{\un\alpha}$ will not close unless the Poisson algebra of
the $H'_{\un\alpha}$ closes. 

One way in which this problem has been approached is to iteratively extend 
the set of operators $H'_\alpha$  until the algebra
closes. Equivalently, new operators in (\ref{X-algebra}) are iteratively introduced,  enhancing the original set
$X_{\un\alpha}$ to a larger set $X_{\bar\alpha}$ of operators until the
closure relation
\be
[X_{\bar\alpha}\,, X_{\bar\beta}] = c_{\bar\alpha\bar\beta}^{\,\,\bar\gamma}X_{\bar\gamma}\,.
\ee
is achieved \cite{Guler92c}.
The corresponding system of partial differential equations will then be integrable.
However, there is no general reason why the solution for the Hamilton principal
function $S$ for this extended first-class algebra should provide the solution for the problem in question.

As an alternative approach to formulate a consistent set of HJ equations,
one can consider \cite{HR} the embedding of the second class system into a first class one characterized by a strongly involutive algebra, following the BFT \cite{BFT} construction. This was not the point of view
to be taken in this paper. It was rather our objective to confront ourselves
directly with the problem of non-integrability, and to point out a procedure
for coping with this problem. The procedure consisted in making a canonical
transformation to a new set of variables in which the second class constraints
become part of the variables, grouped in canonically conjugate pairs. As we illustrated
by a number of examples, the new Hamiltonian is then obtained by setting the constraints strongly equal to zero, and the HJ equations associated with
the constraints are no longer in conflict with each other, nor with this 
new Hamiltonian. The solution for the coordinates obtained
in this way were shown to coincide with the solutions of the
corresponding Lagrange equations of motion.

If the constraints are nonlinear, one is in general not able to perform
a canonical transformation, turning these constraints into canonical pairs
\cite{Senja},
and our analysis as presented here is no longer valid. Investigations
in this directions are presently under way.

\bigskip\noindent
{\bf Acknowledgement}

\bigskip\noindent
One of the authors (K.D.R) would like to thank the Department of Physics
of the University of Stellenbosch for the warm hospitality extended to him
during his stay, as well as Heinz J. Rothe for a useful discussion.

\end{document}